\documentclass[11pt,letterpaper]{article}

\usepackage[superscript]{cite}

\usepackage{hyperref}
\usepackage{float}
\usepackage{pslatex}
\usepackage{amsmath,amssymb}
\usepackage{graphicx}
\usepackage{subcaption}
\usepackage{caption}
\usepackage{color}
\usepackage{url}
\usepackage{todonotes}
\usepackage{mathtools}
\usepackage{stmaryrd}
\usepackage{booktabs}
\usepackage{array}
\usepackage{nicefrac}
\usepackage{amsmath}

\usepackage{url}
\usepackage{authblk}
\usepackage{caption,setspace}
\usepackage{xcolor}
\usepackage{comment}

\usepackage{changepage}
\usepackage{subcaption}
\usepackage{rotating}
\usepackage{color}
\usepackage[letterpaper, margin=0.9in]{geometry}
\title{\textbf{Flexible social inference facilitates targeted \\ social learning when rewards are not observable \vspace{1em}}}
\date{}
\author[1,2\thanks{Correspondence concerning this article should be addressed to Robert D. Hawkins, University of Wisconsin--Madison. E-mail: \texttt{rdhawkins@wisc.edu}. We have released all code, data, and experimental materials required to reproduce the analyses in this manuscript at \url{https://github.com/hawkrobe/emergent-sensing}.}]{Robert D. Hawkins}
\author[3]{Andrew M. Berdahl}
\author[4]{Alex ``Sandy'' Pentland}
\author[5]{\\Joshua B. Tenenbaum}
\author[1,6]{Noah D. Goodman}
\author[5,7]{P. M. Krafft \vspace{1em}}

\affil[1]{Department of Psychology, Stanford University, Stanford, CA, USA}
\affil[2]{Department of Psychology, University of Wisconsin--Madison, Madison, WI, USA}
\affil[3]{School of Aquatic and Fishery Sciences, University of Washington, Seattle, WA, USA}
\affil[4]{Massachusetts Institute of Technology Media Lab, MIT, Cambridge, MA, USA}
\affil[5]{Department of Brain and Cognitive Sciences, MIT, Cambridge, MA, USA}
\affil[6]{Department of Computer Science, Stanford University, Stanford, CA, USA}
\affil[7]{Creative Computing Institute, University of Arts London, London, UK \vspace{2em}}

\begin{document}
\maketitle
\vspace{-4em}

\begin{abstract}

Groups coordinate more effectively when individuals are able to learn from others' successes.
But acquiring such knowledge is not always easy, especially in real-world environments where success is hidden from public view. 
We suggest that social inference capacities may help bridge this gap, allowing individuals to update their beliefs about others' underlying knowledge and success from observable trajectories of behavior.
We compared our social inference model against simpler heuristics in three studies of human behavior in a collective sensing task.
In Experiment 1, we found that average performance improves as a function of group size at a rate greater than predicted by non-inferential models. 
Experiment 2 introduced artificial agents to evaluate how individuals selectively rely on social information. 
Experiment 3 generalized these findings to a more complex reward landscape.
Taken together, our findings provide new insight into the relationship between individual social cognition and the flexibility of collective behavior.
\vspace{4em}
\end{abstract}

\section{Introduction}

Both human and non-human animals make use of social information from others around them.
But not all social information is equally useful. 
Relying on others can be as risky as it is rewarding. 
Social learners must disentangle good advice from bad, and balance the potential benefits of shared wisdom against the risks of being led astray. 
A large body of empirical and computational work in cognitive science and evolutionary ecology has explored specific strategies and heuristics that allow social learning to be most effective \cite{lalandsocial2004,hoppitt2013social,molleman2014consistent,Canteloup2020,RendellFogartyLaland11CognitiveCulture,laland2017darwin}. 
Indiscriminate copying, for example, is not an effective strategy. 
As more individuals rely on imitation, rather than relying on their own independent asocial learning, it becomes increasingly likely that a random target of imitation is using outdated or inaccurate information, degrading the innovation and outcomes of the group \cite{rogersdoes1988}.
For the group to benefit, imitation must be deployed selectively \cite{kameda2003does,boyd1995does,kendal2005trade}, both in choosing the appropriate time to learn from others (``when'' strategies) and choosing the appropriate individuals to learn from (``who'' strategies). 
For example, a ``copy-when-uncertain'' heuristic allows an individual to deploy social learning only when independent learning becomes challenging. 
Or, a ``copy-successful-individuals'' heuristic allows an individual to filter out low-quality social information and target other individuals most likely to increase their own outcomes.

In recent years, the study of social learning has increasingly turned from documenting evidence for isolated strategies or heuristics to investigating  the flexible use of different strategies \cite{heyes2016blackboxing,kendal2018social}. 
Participants in social learning experiments often use hybrid strategies, combining multiple sources of ``who'' and ``when'' information, or deploy different strategies in different contexts \cite{mcelreathbeyond2008,deffner2020dynamic,najar2020actions,toyokawa2019social}.
Thus, it may be useful to view social learning behavior not as the application of an inventory of simple copying rules, but as meaningfully selected social learning behaviours structured by deeper cognitive abilities.
Especially for humans, and some non-human primates, there has been substantial interest in the extent to which social learning relies on abilities like meta-cognition \cite{heyes2016knows} or theory of mind \cite{shafto2012learning} that go beyond pure associative learning \cite{behrens2008associative,heyeswhats2012,heyes2012simple}.
These abilities allow individuals to maintain explicit representations of ``who knows'' and thus concentrate social learning on particularly knowledgeable targets.
Similar capacities have been implicated in organization science as predictors of collective intelligence in small groups \cite{woolley2010evidence,engel2014reading}.

In this paper, we ask whether human social inference abilities may shed light on a puzzle raised by ``who''-strategies like ``copy-successful-individuals.'' 
How is knowledge of success actually acquired when rewards are not directly observable?
Computational simulations \cite{schlag1998imitate,lazer2007network,rendellwhy2010} and human experiments \cite{mason2008propagation,mesoudi2008experimental,mason2012collaborative,derex2013experimental} typically provide individuals in the experiments the ability to directly observe the private information of others (sometimes at a cost for the information). In such experiments, ``success" in the experimental task---i.e., high task performance---is made clearly and unambiguously visible by displaying the scores of other participants, so that determining who is successful in the task does not require inference.
A similar assumption is made in studies of ``transactive memory systems" \cite{wegner1987transactive, peltokorpi2019communication} in the collective intelligence literature, where individuals receive explicit information about ``who knows what.''

However, most real-world situations do not involve such direct and unambiguous social information, posing a challenge for heuristic accounts. 
When information about the success or payoff of an individual's actions is hidden from others, the benefits of selective copying may be negated or even reversed, as the solutions of different individuals cannot be compared \cite{wisdomsocial2013}.
Thus, accounts of selective copying that rely on information about who is successful or knowledgeable must also provide an account of how individuals come to know about others' success or knowledge.
While it is possible that associative learning allows individuals to adopt particular external cues as proxies for private information (e.g. visible health or conspicuous wealth), we suggest that social inference abilities present a more general, flexible alternative. 
Humans continually move between different contexts where ``success" manifests in different observable behaviors: a reliable cue of success in one environment may not be reliable in another.
By inverting a generative model of behavior \cite{jara2016naive,baker2017rational}, individuals can make context-sensitive predictions and flexibly infer the hidden success or knowledge of others.

Such inference abilities have been extensively investigated in cognitive science.
Studies have shown that even young children are able to rapidly infer which partners are more trustworthy and knowledgeable than others, and prefer to learn from those partners \cite{wood2013whom,sobel2013knowledge,poulin2016developmental,mills2016learning}.
Adults also appropriately discount unreliable social information in their decision-making \cite{hawthorne2019reasoning,velez2019integrating,WhalenEtAl18SensitivityToSharedInfo}.
However, this cognitive science literature has largely developed independently from the literature on collective behaviour, and the consequences of these inference abilities for groups remain unclear.
This gap may attributed in part to the significant methodological challenges associated with running real-time, interactive studies with human groups at the necessary scale\cite{almaatouq2021task, almaatouq2022beyond}.
In this work, we bridge these two literatures by examining the behavior of human groups in a large-scale collective sensing experiment.

\begin{figure}[h!]
  \centering
 \includegraphics[width=.99\textwidth]{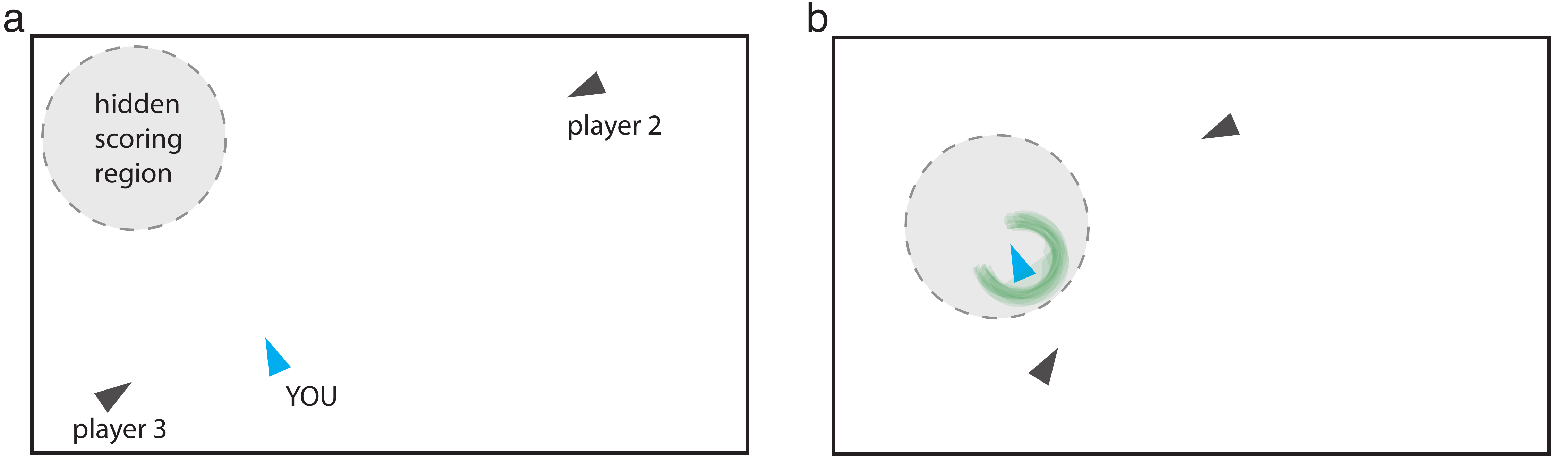}
  \caption{Example states of the collective sensing task used in Experiment 1 and in computational simulations. (a) The hidden scoring "spotlight" region is shown in grey. This spotlight area slowly drifts over time. (b) Participants receive a bonus reward upon entering the region. The halo indicating this bonus was only visible to the participant inside the region, and not to the other participants.\vspace{2em}}
  \label{fig:score}
\end{figure}

Collective sensing tasks are defined by a spatial reward landscape in which multiple interacting agents search for some hidden information about the environment. 
This type of task is a form of collective navigation \cite{berdahl2018collective}, most specifically related to collective foraging  \cite{dechaume2005hidden,goldstone2005knowledge}. 
Unlike foraging, however, the hidden `resource' is not consumable, so there are no competitive dynamics. 
Importantly, while individuals have private information about their own success in the search task, this information is not directly observable by other agents.  
The specific collective sensing design we use builds on an experiment that was designed to study the collective sensing of fish schools \cite{berdahlemergent2013}, as well as from the economics literature on social learning with private information \cite{bikhchandani1998learning,anderson1997information}.

In our experiments, human participants controlled avatars in a virtual world (\autoref{fig:score}).
Each location corresponded to a hidden score value that fluctuated over time.
Participants could continually observe the movements of other individuals but only had access to the score at their own current location. 
Across three experiments, we used this virtual environment to investigate how participants tracked the fluctuating score field as a function of group size (Experiment 1); to carefully evaluate the individual strategies driving collective success (Experiment 2); and to study the effect of increasing environment noise/uncertainty on social learning (Experiment 3).
Taken together, our work suggests that in novel environments where rewards are not directly accessible, even rudimentary forms of social inference can enable targeted social learning that boosts the group's overall performance. These findings emphasize the importance of flexible inference abilities for understanding how people make sense of ambiguous social cues in social learning settings, and further elaborate the centrality of social reasoning to human collective behavior.

\section{Results}

\subsection{Comparing computational models of social learning}

\paragraph{Model overview}

How should individuals weigh ambiguous social cues when they do not have direct knowledge of others' payoffs? 
In this section, we formalize our key hypothesis: when different private information leads to different public (task-specific) behavior, individuals equipped with a rudimentary capacity for social cognition may use observations of visible behavioral trajectories to make flexible inferences about hidden variables.
As an illustrative example, consider a group of students exploring a large food court \cite{baker2017rational}.
One student in the group may reason that people are more likely to sit down longer to enjoy a meal at a food truck they find they like, and move on quickly from food trucks they don't like---perhaps without even finishing their meal.
This kind of reasoning is sometimes called a ``forward model'' or ``generative model'' of social reasoning.
The agent is able to make different social predictions depending on unobservable latent states like beliefs and preferences. 
Critically, a forward model can be inverted to yield predictions about those latent states given observations.
For example, after seeing someone lingering at a particular vendor, our student may infer that this vendor is more likely to sell good food and head over to try it out themselves. 

This idea is naturally formalized in recent probabilistic models of social cognition \cite{jara2016naive,baker2017rational}.
These models propose that inferential reasoning can be understood as (approximate) Bayesian belief-updating using rich probabilistic models of the world. 
We formulate and test a social inference model in this framework, and compare it against alternatives that suppose human behavior in our experiments can be explained by simpler perceptual heuristic strategies that do not employ explicit social inference. 
Under our social inference model, agents do not simply operate on shallow perceptual cues, but are able to update their beliefs about the latent states of other agents (e.g. the private reward they are getting) and use these inferences to guide their own downstream decision-making, similar to related models\cite{perez2011collective,karpas2017information,mann2018collective,mann2020collective}. 
Formally, Bayes' Rule gives the ``posterior probability'' an agent should assign to different (hidden) states of affairs $s$ after making some observations $o$:
$$P(s \mid o) = \frac{P(o \mid s)~P(s)}{P(o)}$$
Intuitively, this equation decomposes the posterior probability into two terms. 
The likelihood term $P(o \mid s)$ represents the probability that $o$ would be observed under different states $s$.
The prior term $P(s)$ represents the background probability of state $s$ in the absence of any additional information.

\paragraph{Task formulation}

As a more concrete domain to explore the consequences of social inferences for collective behavior, we take inspiration from the collective sensing literature \cite{berdahlemergent2013,hein2015evolution}.
In a collective sensing task, a group of agents is placed in a dynamic environment with a shared ``score field'' that gradually shifts over time (see \autoref{fig:score}).
This underlying field determines the scalar reward obtained at each spatial location at each time step in the environment. 
Each agent is only able to observe the private reward $r^t$ they are earning at their current location; they cannot see the rewards available elsewhere in the environment, or the rewards currently being obtained by other agents.
To succeed at this task, agents must balance exploitation (staying in a known area of high reward) with exploration (searching for unknown regions that may provide even higher reward).
Importantly, however, agents have another source of information at their disposable: the public movement trajectories of others, $\{\tau_1^{0:t}, \dots, \tau_{N-1}^{0:t}\}$, where $N$ is the group size and  $\tau_i^{0:t}$ denotes the movement trajectory of agent $i$ from time 0 up to the current time step $t$. 

We hypothesise that people are able to use social expectations about how other agents will (publicly) act in order to back out information about (private) reward at distal regions of the environment. 
Using Bayes' Rule, an agent's beliefs about the (private) reward being obtained by another agent $i$ in the environment, $P(r_i^t \mid \tau_i^{0:t})$, is derived from integrating their background prior beliefs about the overall distribution of rewards in the environment $P(r_i^t)$ and the likelihood that agent $i$ would produce that movement trajectory if they were earning a reward, $P(\tau_i^t \mid r_i^t)$,  relative to the way they would behave if they were not earning a reward. 
To simplify this calculation, we designed the environment to use a binary score field, yielding particularly strong statistical information from behavioral trajectories. 
When an agent gets a non-zero reward, they may expected to slow down or stop relative to the speed they would ordinarily move when earning no reward.
For shorthand, we call this kind of slowing event ``exploiting behavior`` ($\tau_i^t$ = \texttt{exploit}). 
If we assume agents are much more likely to exploit at times when they are receiving reward, conditionally independently of their earlier trajectory, i.e. 
\begin{align*} P(\tau_i^t = \texttt{exploit} \mid \tau_i^{0:t-1}, 
 \,\,\, r_i^t = 1) & = P(\tau_i^t = \texttt{exploit} \mid r_i^t = 1) = 1-\epsilon\hspace{1em}\\ & \gg P(\tau_i^t = \texttt{exploit} \mid r_i^t = 0) \approx 0\end{align*}
then, via Bayes' Rule, an agent's posterior beliefs are dominated by the likelihood term,
$P(r_i^t = 1 \mid \tau_i^t = \texttt{exploit}) \approx  1-\epsilon$.
This equation corresponds to the inference that if an individual stops at some location, it is very likely that there is a reward there (otherwise they would be unlikely to stop).
Before presenting alternative models, we make two subtle but important observations. 
First, this model does not require any intentional signaling or recognition of intentional signaling \cite{torney2011signalling,brown1988social,brown1991food}; the inference is based entirely on spontaneous behavioral trajectories that appear as a consequence of reward-seeking activity. 
Second, this inference is contingent on the details of the local task environment; the visual appearance of successful behavior may differ from situation to situation.
For example, if the score field had different dynamics, or reward was more distributed throughout the landscape, stopping or slowing may be a sub-optimal response and thus an unreliable cue to reward.
The flexibility of social inference is derived from having a general-purpose forward model of how agents with certain goals and information will actually act. 
Neither our simulated agents nor our human participants have had prior experience with the specific interface we use in our experiment, so it is not obvious why ``copy players who stop or slow down'' would be specifically pre-equipped as a generic heuristic applicable across multiple environments.

\paragraph{Simulation and model comparison}

After formulating our social inference model, we compared its predictions against simpler heuristic alternatives by simulating groups of agents following different strategies.
As alternative models, we consider (1) an asocial model, (2) a ``move-to-center" heuristic, and (3) a ``naive copy" heuristic. 
In contrast to our social inference model, an agent implementing the baseline ``asocial'' strategy does not pay any attention at all to other agents and randomly chooses destinations to explore. 
An agent implementing the ``move-to-center'' heuristic similarly sets a destination to explore but is biased toward the centroid of the other agents' positions at the time of their choice, where the degree of bias is a free parameter. 
Finally, an agent implementing the ``naive copy'' heuristic randomly chooses between independently exploring versus indiscriminately choosing a single agent to copy (i.e., choosing uniformly from the set of other agents, without making an inference about their current reward).
These models have one free parameter, the ``independent explore probability'' $\theta_{\textrm{exp}}$, which governs the frequency with which agents set an independent destination as opposed to setting a social heuristic destination.
When $\theta_{\textrm{exp}}=1$, these models are equivalent to the asocial model; when $\theta_{\textrm{exp}} =0$, agents are strongly tethered to one another and end up stuck in a clump.
For the score field determining rewards in our simulations, we used a slowly moving circular ``spotlight'' along paths between randomly chosen locations (\autoref{fig:score}).
This field was hidden from agents, who only had access to the score at their current location.

\begin{figure}[t!]
  \centering
 \includegraphics[width=0.75\textwidth]{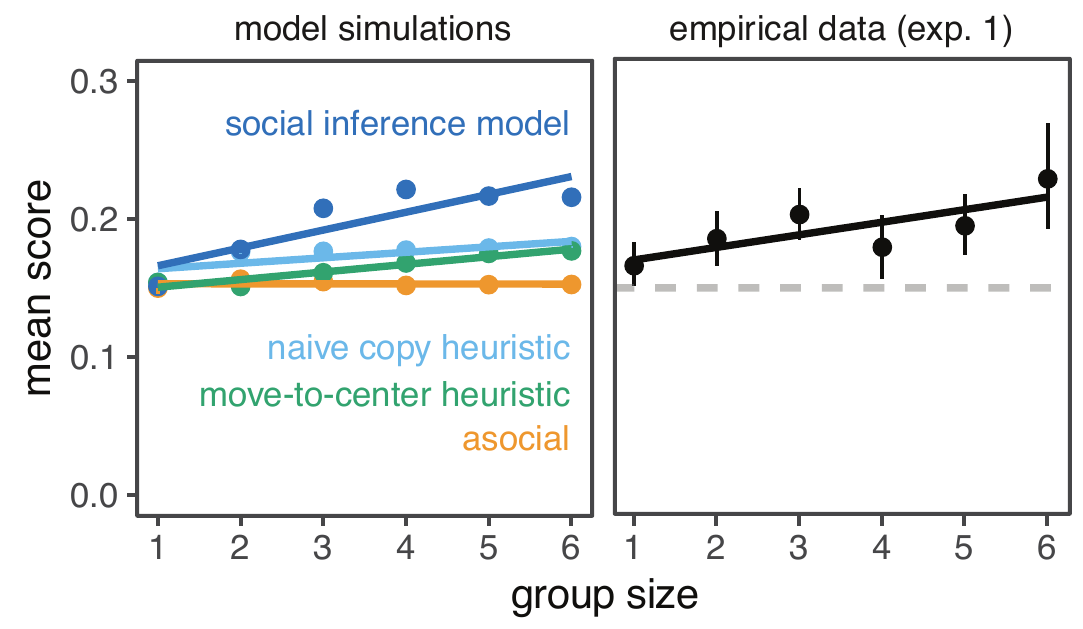}
  \caption{Results of simulations and Experiment 1. (a) Predictions of different models at best-fitting parameter values (50 replicates), (b) Mean performance of human participants in second half of Experiment 1 as a function of group size $(N=682$ participants). Larger groups saw significant gains in performance. Error bars are 95\% bootstrap confidence intervals using unique groups $(N=294)$ as the bootstrap unit. Only the social inference model is able to account for the magnitude of the empirical group size effect.\vspace{2em}}
  \label{fig:exp1_performance}
\end{figure}

We simulated homogeneous groups of different sizes in this environment and measured the average score achieved across the entire group; \autoref{fig:exp1_performance} (left) shows 50 independent replicates under each parameter setting.
``Groups'' of size one were included to establish the baseline level of performance in the absence of social partners.
While we also examined heterogeneous or ``hybrid'' groups, mixing together agents using different strategies, we found that performance in these groups was strictly dominated by the proportion of agents using the social inference strategy, hence we have omitted these simulations for clarity.
Examining the results, we found that groups containing individuals with the capacity to make social inferences performed significantly better as a function of group size. 
These improvements were greater than those observed in alternative models.
First, as expected, the asocial model predicts no increase in performance as a function of group size: agents entirely ignore one another and remain at the individual baseline  ($m\approx 0.15$). 
Second, the heuristic models predict a small improvement over the asocial baseline, but even at the best-performing parameter values, they quickly hit a performance ceiling ($m<0.2$ for groups up to size six; this threshold is only exceeded in much larger groups of size 16 to 32).
The best-performing ``explore'' parameter values $\theta$ for the ``move-to-center'' model and ``naive copy'' model were $\hat{\theta}_{\textrm{exp}}=0.3$ and $\hat{ \theta}_{\textrm{exp}}=0.1$, respectively.
Meanwhile, the social inference model already yields substantially improved performance for small groups ($m>0.2$ already at size 3).
Thus, in purely qualitative terms, groups with the capacity to infer latent reward information from external behavior obtain larger benefits of group size above and beyond those capable of being predicted by simpler heuristics under any parameter regime.
However, it is not clear a priori which of these strategies best explains the behavior of real human groups.

\subsection{Experiment 1: Collective sensing across group sizes}

To evaluate the predictions of these different models, we designed an interactive collective sensing environment where we could examine how groups of human participants $(N=682)$ behave under the same conditions as the simulated agents we considered in the previous section.
Participants were connected into groups of size 1 through 6 over the web and controlled avatars by clicking and using two keyboard keys. 
Their avatars automatically moved forward, and clicking within the playing area instantly oriented the avatar to move toward the clicked location (visually marked with a cross). 
Participants could hold the ``a" key to accelerate or hold down the ``s" key to stop.  
We used the same ``spotlight'' score fields as in our simulations (i.e.,\ a reward of 1 inside the circular scoring region and 0 everywhere else), and showed participants binary feedback about their current score.
Critically, this feedback was only visible to the participant controlling that avatar; participants did not directly observe whether other participants were in the scoring region. 
They could only see the spatial location and orientation of other participants, which were updated in real time (see Supplementary Figure~1 for screenshots).

We hypothesized that individuals in larger groups would be able to achieve significantly higher scores on average than individuals in smaller groups, as predicted by the social inference model. 
To test this hypothesis, we examined the average performance of each group across the 5-minute session, although we were primarily interested in the second half of the session, after participants had gotten used to the user interface. 
We constructed a linear mixed-effects regression predicting each individual participant's average score, including fixed effects of the time window (categorical: first vs. second half), the continuous number of participants in their environment (integer one through six, centered), and their interaction.
We included the maximal random effects structure that converged: intercepts and time window effects for each group ID, as well as intercepts and group size effects for each of the four underlying score fields.
First, we examined our key prediction about the simple effect of group size during the second half. 
Average performance was significantly higher for larger groups, $t(6.6) = 2.9$, $p = 0.024$, $b =0.87$,~95\% CI $= [0.26, 1.44]$.
The smallest groups of size 1 earned a second-half score of $m=0.17$,~95\% \textrm{CI} $= [0.15, 0.18]$, compared to a score of $m=0.23$,~95\% \textrm{CI} $= [0.20, 0.25]$for the largest groups of size 6 (see \autoref{fig:exp1_performance}, right).
Second, we found a large main effect of task practice, confirming the need to examine each phase of the task separately. 
Scores were significantly higher on the second half of the session for participants in all group sizes, $t(205.7)= 9.5$, $p < 0.001$, $b = 4.3$, 95\% \textrm{CI} $=[3.4, 5.1]$.
Finally, we observed a significant interaction, $t(198.6)=2.1, p=0.039$, $b=-0.54$, 95\% CI $=[-1.1 -0.03]$, suggesting that performance on the first half was more similar across group sizes, as expected if all groups required equal practice with the interface.

Critically, the magnitude of these performance gains in human groups exceeded the bounds of the two heuristic models in our simulations even at their best-performing parameter values; only the social inference model was capable of explaining the empirical group size effect observed in groups of human participants.
To validate the assumed difference between the likelihood of ``exploiting'' trajectories in the presence vs. absence of reward, we computed the average movement speed for each participant as a function of their current background score.
Consistent with the assumptions of the model, we found that participants indeed move about half as fast when reward is present (2.6 pixels per tick) as they do when reward is absent (5.1 pixels per tick;  $t(1360)=38.9$, $p < 0.001$, $\hat{d}=2.5$; 95\% CI $= [2.37, 2.62],$ in a paired test). 
The size of this effect was roughly constant across all group sizes (see Supplementary Figure~2).
Although we have focused on the qualitative differences between models at the level of group performance, rather than quantitative fits to individual agent behavior, we note that a moderate amount of noise ($\epsilon = 0.15$) must be introduced to the social inference model for an accurate quantitative fit, possibly reflecting human limitations on attention and motor control that lead to deviations from perfect exploitation or copying. 
An optimally executed social inference model exceeds human performance.
This performance gap is also related to our motivation for comparing model performance on second-half scores: our idealized simulated agents do not face the same learning curve as human participants with respect to aspects of the task such as motor control and instruction comprehension, which we are not interested in explicitly modeling in the current work.
Extending our model to a more granular level of analysis, including human attention patterns, is an exciting direction for future work.

\subsection{Experiment 2: Evaluating copying strategies}

\begin{figure}[t!]
  \centering
 \includegraphics[width=1\textwidth]{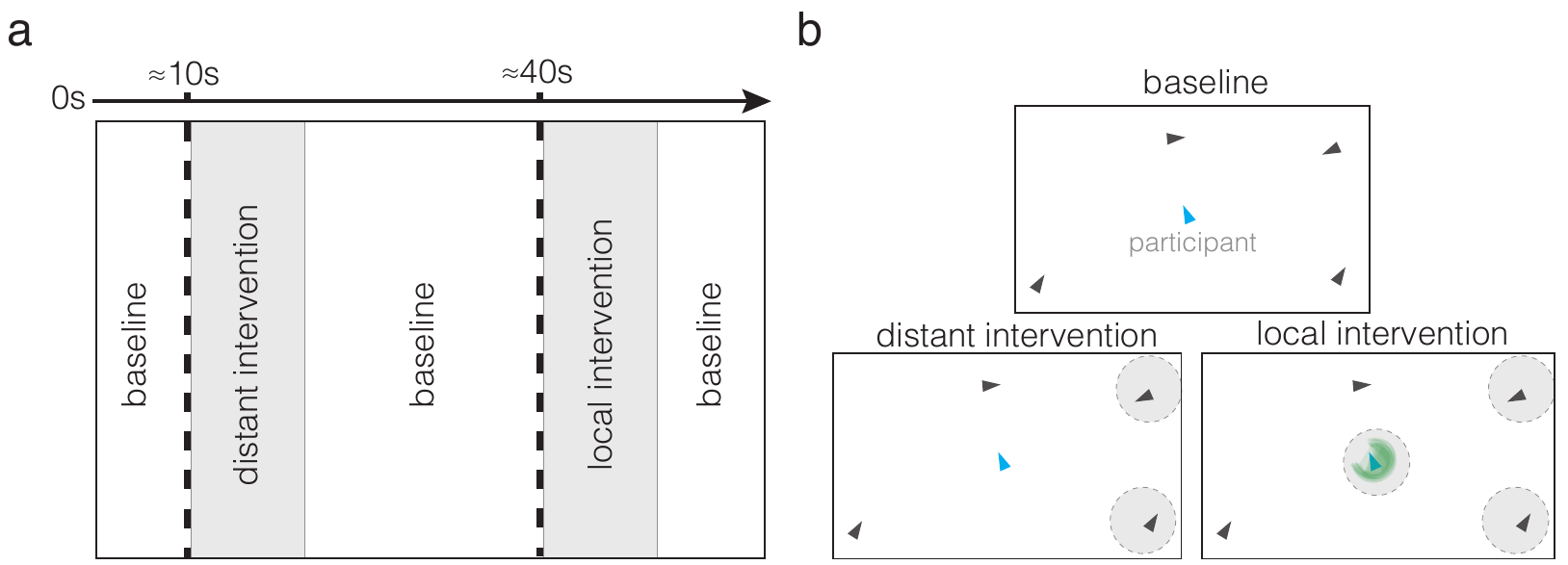}
  \caption{Design of Experiment 2. (a) The timeline of the test round involves a baseline condition with no score field, and two interventions, presented in randomized order, on the score field beginning at approximately 10 seconds and 40 seconds. (b) These interventions manipulate the location of the score field to ensure the participant is receiving high reward or not, respectively, while a subset of other agents are receiving high reward. \vspace{2em}}
  \label{fig:exp2_design}
\end{figure}

What cognitive abilities allowed participants in Experiment 1 to benefit from social learning even when the payoff information of other individuals is not directly observable?
We hypothesized that human behavior in this environment is driven by a combination of two underlying strategies: (1) independent exploration with opportunistic exploitation and (2) selective, targeted copying based on social inferences about success. 
These strategies rely on the ability to infer ``who knows'' about high-scoring locations based on outward behavioral traces (e.g. slowing down or stopping in a region) and also to inhibit social influence and explore independently when copying is inappropriate.
The design of Experiment 1 made it challenging to disentangle these strategies.
For example, we could analyze participant clicks to detect signatures of selective copying, but because there was a unique `spotlight' at each point in time, different copying strategies would be confounded: participants who were already obtaining reward and trying to stay inside the spotlight were, by necessity, clicking close to other participants who were obtaining reward, even if they were not intentionally copying them.

For our second set of experiments, then, we designed a sequence of controlled scenarios that are more diagnostic for testing the use of these different strategies.
Rather than placing participants in groups with other humans, we placed them into a group of artificial agents (bots) that we designed and controlled to follow specific behaviors. 
We also casually manipulated the score field to estimate the probability of copying different agents under different conditions.
As in our first experiment, participants were given control of an avatar to explore a virtual environment and were rewarded based on their location according to a hidden score field. 
The interface and controls were the same as in Experiment 1. 
Instead of a single 5-minute session, however, we designed a sequence of shorter scenarios that were more informative for distinguishing between several different strategies that could be used in the game.

\begin{figure}[t!]
    \centering\includegraphics[width=0.5\linewidth]{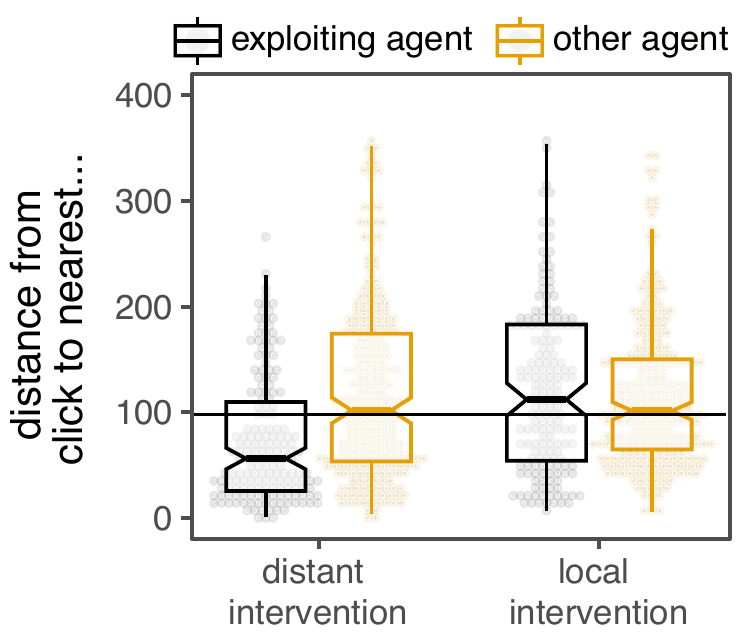}
    \caption{Results of Experiment 2. Participants selectively copy other agents who appear to be exploiting, but only when they themselves are not receiving reward (i.e. in the distant intervention condition). We operationalized copying in terms of the spatial distance from the click to the (nearest) location of another agent. Dark horizontal line represents the average value in the baseline condition, where no agents were exploiting. The top and bottom of box represent first and third quartiles, respectively; center line represents median, and whiskers extend to the most extreme values 1.5 times the inter-quartile range. Notch represents bootstrapped 95\% CI for median $(N=28$ participants). \vspace{2em}}
    \label{fig:proximity}
\end{figure}

After four one-minute practice rounds, where no other agents were present, participants were placed in two one-minute test rounds that were the focus of our analyses. 
In one of the two test rounds, no other agents were visible (non-social condition), and in the other there were four visible bots in the environment (social condition). 
Each of these rounds contained two randomized interventions on the score field. 
Most of the session was spent in a baseline condition, where reward was exactly zero at every location in the task environment. 
Around the ten second mark and the forty second mark in each round, we  introduced high-scoring regions into the game (see \autoref{fig:exp2_design}a).
In one condition (distant intervention), we placed these regions directly on top of two bots.
In the other condition (local intervention), we also placed a high scoring region at the participant's location, wherever they happened to be at the time, such that they automatically received a high score for the roughly ten-second duration that the high scoring regions were present (see \autoref{fig:exp2_design}b). 
We used exactly the same underlying score field in the social and non-social sessions so that we could compare clicks in the non-social environment to the (hypothetical) locations of bots in the social environment.

For conceptual clarity about our predictions, it is helpful to define three broad `states' an agent may occupy: exploring, exploiting, and copying \cite{rendellwhy2010}.
We define exploiting as selecting an action that maximizes the expected score given the individual's current knowledge of the environment, i.e. staying close to a known location of the spotlight. 
We define copying as forward motion, sometimes accelerated, toward the location of another agent. 
We define exploring as selecting an action that has an unknown outcome, as when moving to a region without other agents. 
In this environment, exploiting, exploring, and copying behavior were associated with distinct and recognizable movements.
The social inference model can thus be operationalized as selective deployment of these three states: exploiting when one is in a high-scoring region, copying when another agent is inferred to be receiving a high score, based on their outward behavioral trajectory, and exploring otherwise.
%We thus predict that participants will selectively copy stopped or slowed agents (i.e. agents inferred to be at high-reward locations) when they themselves are not receiving reward (as in the distant intervention condition), but will more strongly inhibit social influence while exploiting (as in the local intervention condition).
%We tested this prediction by examining which agents participants choose to copy.

Because clicking near another agent set their destination at that target's location, and success is based on spatial location, we operationalized copying via the proximity of each click to other agents.
To examine which agents participants selectively copy, we compared the distance from the click to the nearest agent who is stopped (exploiting), and to the nearest agent who is not stopped.
We then compared copying rates across the two randomized intervention period.
We predicted that participants would inhibit copying to a greater extent in the local intervention condition, when they automatically received a score in their current location, relative to the distant intervention condition, when the score field was only placed on top of other (simulated) agents, and participants were not themselves receiving any reward.
To test this prediction, we constructed a mixed-effects regression model predicting the proximity of each click to the nearest agent as a function of experimental condition (local vs. distant), visible behavior (exploiting vs. not exploiting), and their interaction.
We included the maximal random effects supported by our within-participant design, allowing random intercepts, main effects, and interactions at the participant-level.

We found a significant interaction, $t(25.5)= 2.5$, $p = 0.02$ $b = 47.6$, 95\% CI $= [7.5, 85.4]$, indicating a selective preference for copying other exploiting agents only in the distant intervention condition when the participant was not themselves receiving a reward (see \autoref{fig:proximity}). 
To control for the possibility that this result is a product of generic biases in the spatial pattern of clicks, rather than the use of social information, we conducted the same analysis on clicks in the non-social condition, where no artificial agents were visible but the underlying score field dynamics were the same. 
In other words, this condition allows us to examine the proximity of clicks to where other agents would have been.
We found no significant interaction in this condition; numerically, it was in the opposite direction, $~t(19.9)=1.4$, $p=0.18$, $b=30.1$, 95\% CI $=[-75.9, 17.3]$.
Note that degrees of freedom are lower because not all participants clicked in the corresponding windows.
Our initial power analysis did not take into account the need to estimate the stronger three-way interaction needed to test the difference between these interactions, hence better estimating the the baseline variability of clicks in non-social environments is likely to be a fruitful target for future work using a more highly-powered sample.
Taken together, these results suggest that human participants selectively copy exploiting bots but only when they are not themselves receiving high scores.

\subsection{Experiment 3: Generalizing to more complex environments}

\begin{figure}[t!]
  \centering\includegraphics[width=0.95\textwidth]{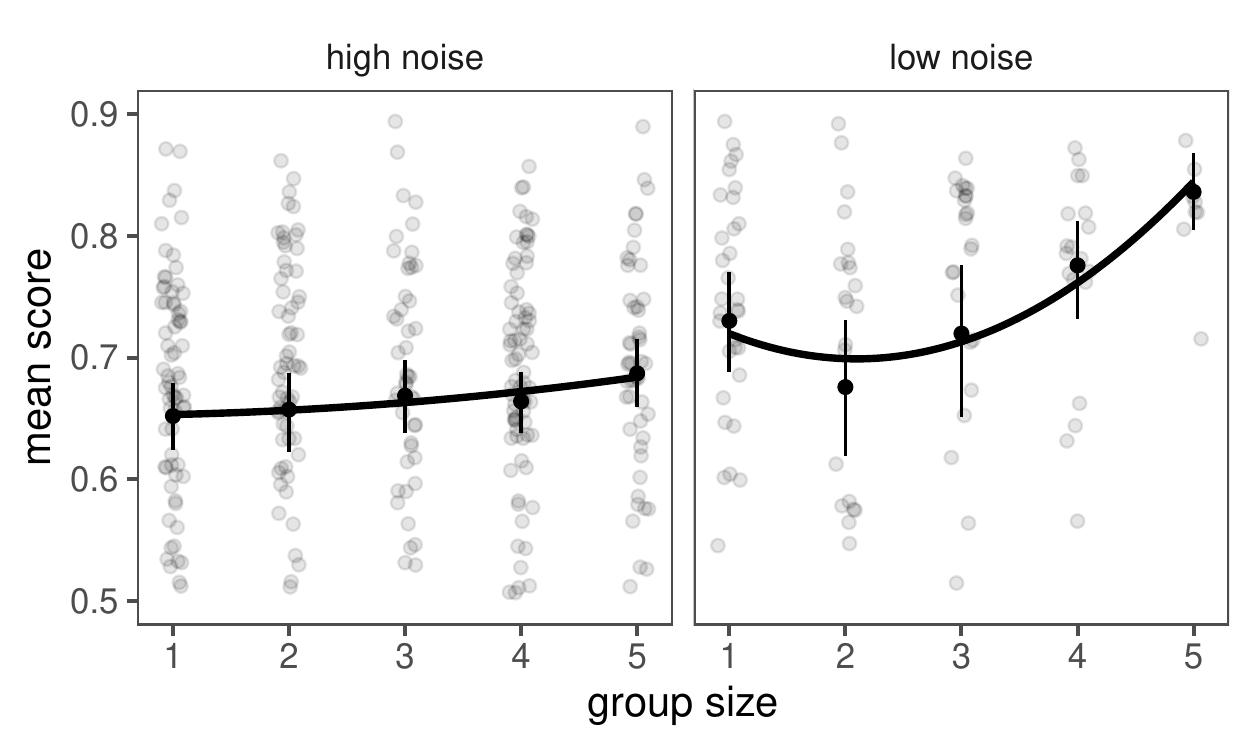}
  \caption{Results of Experiment 3. Mean performance as a function of group size under different noise conditions ($N=454$ participants in $224$ unique groups).  Error bars are 95\% bootstrap confidence intervals using the group as the bootstrap unit. \vspace{2em}}
  \label{fig:performance}
\end{figure}

To generalize our understanding of these findings to more complex environments, we conducted a final experiment using the specific conditions and materials designed by Berdahl et al. \cite{berdahlemergent2013} to examine collective sensing in fish, our prior experiments having used simplifications of this environment.
These environments are substantially more complex than the binary spotlight and border environments we used in Experiments 1 and 2.
They require individuals to use continuous gradients to navigate noisy and fluctuating score fields.
We manipulated the level of noise across different groups, predicting that the cognitive abilities discussed in the previous sections may be less reliable under noisier conditions. To test that the social learning strategies identified in the previous experiments also generalize to different external behavioral signatures, we also modified several other aspects of the experiment interface, including the movement controls: participants used the left and right arrows to change direction instead of clicking at a destination.
This small interface change created a different behavioral cue of success (spinning in place rather than stopping or slowing), which should be just as effective under a social inference account as the behavioral cues in the previous section.
Our analyses focus on two primary questions: (1) how does the introduction of a noisier environment affect average performance, and (2) how do selective social learning strategies play out in such an environment, when inferences about the success of other individuals may be less reliable?

We begin by analyzing patterns of average performance across groups of different sizes and across the different noise conditions.
As our measure of performance, we computed the average score obtained by each participant in the second half of the experiment (i.e. scores obtained after the opening 2.5 minutes of play, for comparability with Experiment 1).
We conducted a mixed-effects regression model including linear and quadratic fixed effects of group size (integers 1 through 5), and noise condition (effect coded: `low' vs. `high'), as well as their interaction. 
We included random intercepts for each group (i.e. controlling for dependence between participants in the same group) and for each of the four distinct score field in each noise condition (i.e. controlling for the possibility that some randomly generated score fields were more difficult than others).
Both main effects were significant: all else being equal, scores tended to increase with group size, $t(81.6)=2.9,~p =0.005,$ $b=0.53,$ 95\% CI $= [0.17, 0.88]$ and were higher overall in the low-noise condition than the high-noise condition, $~t(9.2)=3.6,~p=0.006, b=-0.04,$ 95\% CI $=[-0.06, -0.01]$.
We also found a weak but significant interaction between noise condition and group size, $t(81.6)=2.3,~p=0.02$, $b=0.43,$ 95\% CI $=[-0.78, -0.07]$, indicating a stronger effect of group size in the low-noise condition than the high-noise condition (see \autoref{fig:performance}).
Due to relatively low power to detect the interaction at the group-level unit of analysis, especially given imbalances in sample sizes across noise conditions, we are cautious about overinterpreting this effect but believe it merits further investigation in future work.
Overall, these results indicate an important role of the environment in group success: under low noise, larger groups perform systemically better than smaller groups, similar to the effect found in Experiment 1, yet this advantage appears to be somewhat weakened under high noise.

\begin{figure}[t!]
  \centering
 \includegraphics[width=0.6\textwidth]{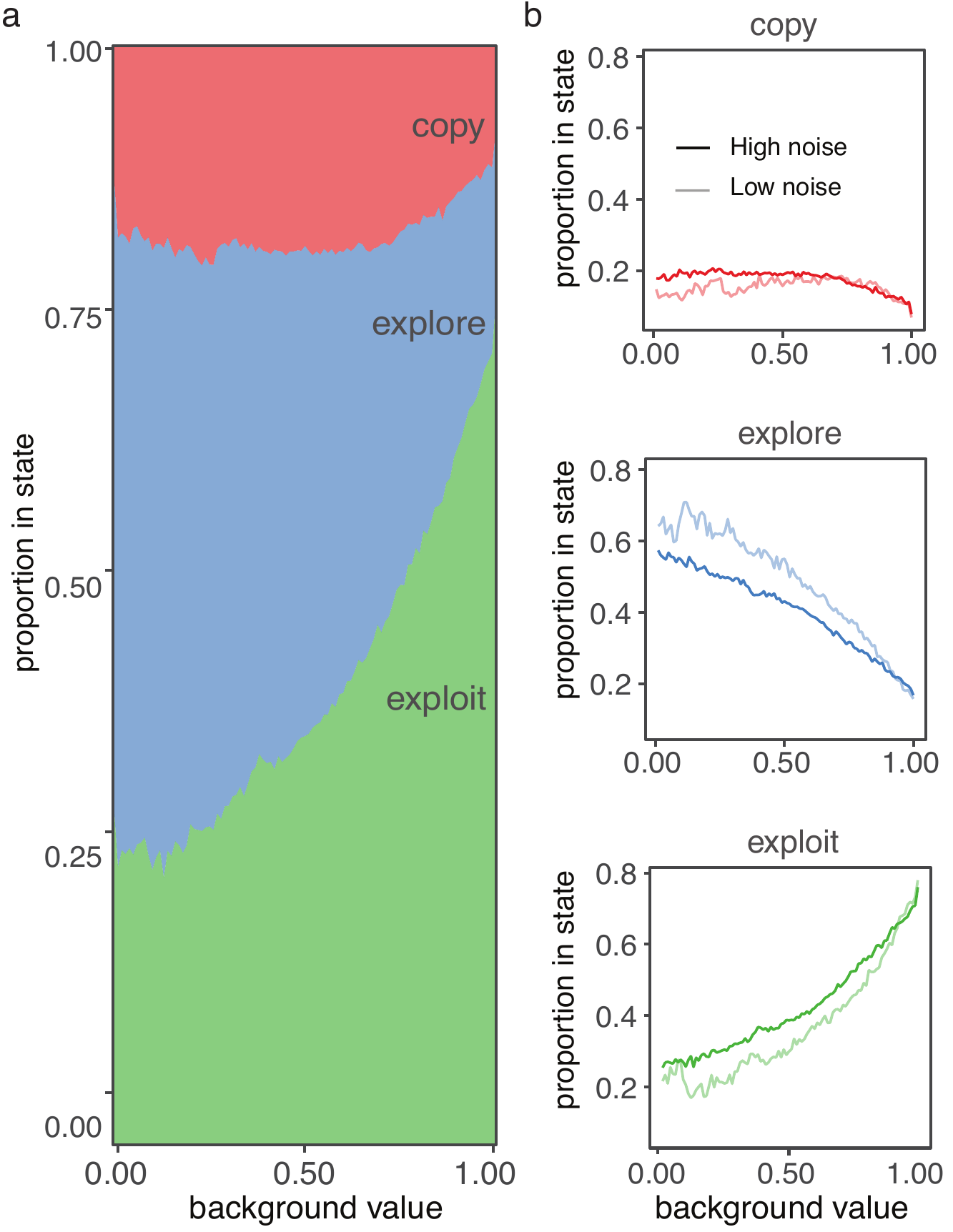}
  \caption{State-based analysis of behavior in Experiment 3. (a) The probability of an individual being in a particular behavioral state as a function of the individual's score, combined across both conditions. (b) Participants tended to begin exploiting at lower background values in the high-noise condition, leading to less copying and exploration. \vspace{2em}}
  \label{fig:states}
\end{figure}

In order to understand the mechanisms that may have contributed to effects of noise on average performance, we more closely analyzed the underlying behavior of the participants in our games.
While we relied on click data as a useful measure of copying in Experiment 2, this measure was no longer available under the new control interface.
Instead, we used a simple state-based coarse-graining of participant trajectories based on their keyboard actions.
We automatically coded the state of each participant at each point in time --- exploring, exploiting, or copying --- using a simple set of criteria (see Methods for details; an annotated example is shown in Supplementary Figure~3, and in the Supplementary Videos).
All of these criteria depended only on public information that was observable to the participant in the game (i.e., the state does not depend directly on the hidden scores of other individuals). 
Thus, while this classification is purely intended for analyzing and interpreting behavioral patterns (distinct from our computational model), we can use these states as proxies for what other participants might in principle be able to infer from external cues.
Additionally, because the coding of these states did not depend in any way on actual score values, we can meaningfully quantify the relationship between state and performance.

We proceeded to use these state classifications to analyze the behavioral strategies used by participants.
First, in line with our earlier findings that participants inhibit copying when they find themselves in high-scoring regions, we predicted that the probability of the exploiting state would increase as participants receive higher scores.  
To test this prediction, we constructed a logistic mixed-effects regression model predicting the probability that each individual is in the `exploiting' state at each time step. 
We included fixed effects of their current background score and noise condition, as well as their interaction.
We also included random intercepts for each group and score field.
First, we found a strong main effect of the current score: regardless of noise condition, participants are significantly more likely to exploit in higher scoring locations than in lower scoring locations, $z=311,~p<0.001,~b=3.22,~$ 95\% CI $=[3.19, 3.24]$ (see \autoref{fig:states}a).
Selective exploiting is clearly adaptive, as participants will tend to remain in high scoring regions but quickly move away from low scoring regions either by exploring independently or by copying other individuals.
At the same time, strategies differed dramatically across noise conditions: we found a significant interaction between condition and score value, $z=-33.9,~p<0.001$, $b=-0.35$, 95\% CI $=[-0.37, -0.33]$, indicating that participants in the high-noise condition begin to exploit at lower  score values (see \autoref{fig:states}b). 
Considering the greatly increased variability both spatially and temporally in the high noise condition, participants may expect a smaller range of achievable scores.
Similar regressions predicting the probability of `copying' and `exploring' states found that participants were also less likely to be exploring, $z=27,~p<0.001$, $b=0.26$, 95\% CI $= [0.24, 0.28]$ and more likely to be copying, $z=5.9,~p<0.001$, $b=0.07$, 95\% CI $= [0.05, 0.10]$ at lower score values in the high-noise condition.

A lower threshold for exploitation may also help explain gaps in average performance across noise conditions. 
First, a willingness to exploit at lower point values may, by definition, lead to lower overall performance. 
Second, it may make copying less effective, preventing social learning mechanisms from improving performance in larger groups.
That is, if participants are willing to exploit at lower background score values, then external cues of exploiting (i.e. ``spinning'' behavior) will provide statistically weaker evidence of underlying success.
To test this hypothesis, we identified all of the events in our dataset where one participant copied another and measured the current score of the target of copying.
We found that targets of copying tend to be in lower scoring regions in the high-noise condition, $d=0.08,~t=4.02,~p<0.001$.
These results clarify the interaction between human social learning strategies and environmental conditions, and raise interesting questions about the robustness of social inference based copying.

\section{Methods}

This research was reviewed and approved by the MIT Committee on the Use of Humans as Experimental Subjects (COUHES), Protocol \#1509172301, as well as the Oxford Internet Institute Departmental Research Ethics Committee (DREC), CUREC 1A Research Ethics Approval Ref Number SSH\_OII\_CIA\_20\_002.

\subsection{Experiment 1: Manipulating group size}

\paragraph{Participants}
We recruited 781 unique participants from Amazon Mechanical Turk and assigned them to one of 322 unique groups for an interactive web experiment \cite{hawkinsconducting2014}. 
All participants were from the United States. 
A number of these participants did not make it past the instructions or comprehension quiz, leaving only 304 unique groups (738 participants) that reached the outset of the task. 
To minimize potential bias from differential dropout rates, we coded group size using the number of players present at the outset of the second half, when we began measuring performance. 
56 of these participants disconnected at some point during the first half of the game due to inactivity or latency, leaving second-half data from 682 unique participants in 294 complete games (10 games terminated early because no participants were remaining).
We calculated the average score across all participants assigned to a group for the time that they were present (including those who dropped out later in this second half).
However, results were qualitatively similar for other exclusion choices. 
We paid participants 75 cents for completing our instructions and comprehension checks, as well as a bonus up to \$1.25 during the five minutes of gameplay. Each ``point'' in the game corresponded to \$0.01 of bonus. Each participant was also paid 15 cents per minute for any time spent in the waiting room, minus any time that participant spent moving into a wall.  These numbers were chosen so that the participants were expected to receive at least a wage of \$9 per hour for the totality of their time active in the experiment.

\paragraph{Stimuli}

The virtual game environment measured 480 pixels in width and 285 pixels in height.
Avatars were represented by triangles that were 10 pixels in length and 5 pixels in width, rotated to the direction the avatar is facing. 
The avatars automatically moved forward at a constant velocity of 17 pixels per second if no buttons were pressed, but instantaneously increased to a constant velocity of 57 pixels per second for the duration of time that the ``a" key was held down and decreased to 0 pixels per second for the duration of time the "s" key was held down. 
Locations were updated every 125 milliseconds.
As soon as an avatar entered the ``spotlight'' reward region, it was surrounded by a salient sparkling halo and the border of the playing area turned green. 
To discourage inactivity, participants also received 2/3 of a point for each second they were actively participating in the game.
For any moment when an avatar was touching a wall, we displayed a large warning message and set the participant's current score to zero so that they stopped accumulating points.
We generated score fields by first initializing a circular region with a diameter of 50 pixels at a random location on the playing area. 
Inside this region, the score was set to 1.
Outside this region, the score was set to 0.
We then moved this region along a straight line to a randomly chosen target location within the playing area at a speed of 17 pixels per second.
Once it reached this location, we selected another target location, and repeated the process for the duration of the 5-minute session.
We pre-generated four unique score fields in this way, and randomly assigned groups to one of these fields.

\paragraph{Procedure}

After agreeing to participate in our experiment, participants were presented with a set of instructions describing the mechanics of the game, using a cover story framing the game as a search for the ``magical bonus region". 
The participants were informed about the dynamics of the underlying score field and also explicitly informed that ``There is no competition between players; the magical region is not consumed by players. It simply changes location over time." 
Participants were not explicitly instructed or suggested to cooperate or coordinate with each other.
After successfully completing a comprehension test, participants were redirected to a waiting room.
Each waiting room was assigned a group size between 1 and 6, and the game began as soon as the target number of participants was reached, or after 5 minutes of waiting, whichever came first.
While in the waiting room, participants could familiarize themselves with the controls of the game.  
Participants were not shown any score in the waiting room unless the participant was against a wall, in which case the border of the playing region would turn red and a warning appeared on screen.  
All participants spent at least one minute in the waiting room to help ensure familiarity with the controls before starting the game. 
Participants then played a single continuous game lasting for 5 minutes, and were paid a bonus through Amazon Mechanical Turk proportional to the total score they (individually) accrued. 
Both in the waiting room and the actual game, participants were removed for inactivity if we detected that they had switched to another browser tab for more than 30 seconds total throughout the game or if the participant's avatar was moving into a wall for 30 consecutive seconds.  
We also removed participants if their ping response latencies were greater than 125ms for more than 75 seconds in total throughout the game.  
To minimize disruption of large groups, we allowed multi-participant games to continue after a participant disconnected or was removed, as long as at one or more participants remained.

\subsection{Experiment 2: Manipulating the behavior of other agents in micro-scenarios}

\paragraph{Participants.}

We recruited 28 unique participants from Amazon Mechanical Turk. 
All participants were from the United States.
A smaller number of participants was required relative to the other experiments, as Experiment 2 consisted only of single-participant conditions rather than interactive groups of multiple participants. 
All manipulations were conducted within-participant, and a preliminary power analysis showed sufficient power to measure the two-way interaction of interest at this sample size.

\paragraph{Stimuli \& procedure.}

To acclimate participants to the task environment, each game began with four one-minute long practice rounds. 
In the first and third practice rounds, the score field was visible to the participant so they could observe its dynamics.
In the second and fourth practice rounds, the score field was invisible to the participants, as in Experiment 1. 
Additionally, we randomized participants into two different groups, who practiced with different score field dynamics. 
In a ``wall-following'' pattern, the high scoring region moved contiguously along the walls of the playing area. 
In a ``random-walk'' pattern, the high scoring region slowly drifted, as in Experiment 1, from one random location to another within the playing region.
Because we did not observe substantial differences in participant behavior depending on the score field dynamics observed during the practice phase, we collapsed over this factor in our analyses.

Bots followed a simple selective copying algorithm. 
They were programmed to immediately stop upon entering a high-scoring area.
If a bot in the environment was stopped, they copied the stopped bot. Otherwise, they explored non-socially.
The wall-following bots only copied other wall-following bots, and the bots in the center region similarly only copied each other.
Bots were not responsive to the participant's behavior, only to each other.
In the non-social round, we simulated where the same bots would be, so that the distribution of score field positions was held constant across the two conditions. 
The score field manipulation was triggered for the bots approximately two seconds after it was triggered for the participant in the local intervention condition. 
We offset the trigger time in order to ensure that participants were already aware of their own score before observing any reward-related bot behavior.

The within-game interventions were implemented for bots as follows.
First, in the baseline condition, when there was no score field and all bots were randomly exploring, two were randomly exploring along walls (in association with the wall score field) and two were exploring the center region (associated with the random walk score field). 
For the interventions, we superimposed the wall-following and random-walk score field patterns to create a bi-modal dynamic score field.
One high-scoring region was centered on a wall-following bot and one high scoring-region was centered on a bot in the center region. 
We randomized both the order of the social and non-social micro-sessions and the order in which the distant and local interventions appeared within each session.

\subsection{Experiment 3: Manipulating noise in the environment}

\paragraph{Participants}

We recruited 515 unique participants from Amazon Mechanical Turk to participate in our experiment.
All participants were from the United States.
After excluding 61 participants who were inactive or disconnected in the first 2.5 minutes of the game (prior to the window used for our performance analyses), we were left with data from 454 participants in 224 complete sessions.
116 individuals (63 groups) were in the low noise condition and 338 individuals (161 groups) were in the high noise condition. 
These groups ranged in size from one to six individuals.  
Since only one group of size six completed the task without disconnections in the low-noise condition, we ignored this group in our analysis.

\paragraph{Stimuli and Procedure}

The primary change from Experiments 1 and 2 was replacing the binary score field with a more complex, gradient score landscape.
These more complex fields were generated using the method reported by Berdahl et al. \cite{berdahlemergent2013}. 
We began with the same randomly moving ``spotlight'' as before, which was also the basis for the Berdahl et al. fields.
However, we then combined the spotlight with a field of spatially correlated, temporally varying noise. 
By manipulating the proportional weighting of the noise field and the spotlight, we generated two different conditions. 
In the low noise condition, the spotlight was weighted strongly compared to the noise field (10\% noise), with the noise field providing minor background variation (see Supplementary Figure~4, left). 
In the high noise condition, the weighting of the noise field was increased (25\% noise), providing more extreme fluctuation outside of the spotlight (see Supplementary Figure~4, right).
To decrease variability and increase statistical power, we generated only four distinct score fields per noise level, so multiple groups experienced the same fields.  

In addition to these more complex score fields, we made several adjustments to the interface.
First, rather than showing their current score as binary---a glowing halo around the participant when inside the spotlight---their score was presented as a percentage at the top of the playing area (see Supplementary Figure~5 for a screenshot).
Second, rather than clicking to change direction, participants controlled their avatars using their keyboard. 
The left and right arrow keys were used to turn (at a rate of $40^\circ$ per second) and the
spacebar was used to accelerate. 
Unlike before, we did not provide a mechanism to stop completely.  
Given the closer relation to the design used by Berdahl et al. \cite{berdahlemergent2013} in this experiment, it is also relevant that the baseline speeds of the avatars and the playing area dimensions ($480 \times 285$) throughout all of our experiments were chosen to match those reported in animal work; for this experiment, we additionally matched their total task length of six minutes (as opposed to five minutes in Experiment 1).
The procedure was otherwise identical to Experiment 1.

\paragraph{Automated state classification}

Our criteria for classifying agents into one of the three states are as follows.
\begin{itemize}
\item \emph{Exploiting} behavior was not trivial for participants since the avatars always move at least at a slow constant velocity. 
Unlike in the previous experiments, where the ``s'' could could be pressed to stop in place, a participant in Experiment 3 could either slowly meander around a particular location or persistently hold down one of the arrow keys while moving at a slow speed, which creates a relatively tight circular motion around a particular location.  
We call this second activity ``spinning'' because of its distinctive appearance.  
We classify a participant as exploiting if the participant is spinning for 1 second, or if the participant moves at a slow speed for 3 seconds and has not traveled more than two thirds of the possible distance that the participant could have traveled in that time (i.e. participants who are meandering around a small area and may not yet have discovered how to spin).
\item \emph{Copying} behavior is also more difficult to identify in this more complex environment, but is characterized by directed movements towards other participants. 
We thus classify a participant as copying if they move at the fast speed in a straight line towards any particular other participant for a 500ms window.
We consider a participant to be moving towards another participant if the second participant is within a $60^\circ$ on either side of the first participant's straight-line trajectory. It is possible that this coding scheme under-estimates the overall rates of what we might still want to call `copying' behaviors: for example, it does not capture more graded biases toward the group centroid. Because such under-estimation would be constant across background score values, we do not expect it to affect the comparisons of interest.
\item We classify behavior as \emph{exploring} if the participant is neither exploiting nor copying. Thus, a participant will be classified as exploring if that participant is either moving slowly but not staying in the same general location or if the participant is moving quickly but not towards any particular person.
\end{itemize}

\section*{Data Availability}

All data available at \url{https://github.com/hawkrobe/emergent-sensing}.

\section*{Code Availability}

All experiment and analysis code available at \url{https://github.com/hawkrobe/emergent-sensing}.

\section*{Acknowledgments}

A preliminary version of our work reporting Experiment 3 appeared at the 2015 Annual Conference of the Cognitive Science Society.
This work was supported by the Center for Minds, Brains and Machines (CBMM), funded by NSF STC award CCF-1231216, as well as NSF Graduate Research Fellowships under Grant No. 1122374 to PK and Grant No. DGE-114747 to RDH. Any opinion, findings, and conclusions or recommendations expressed in this material are those of the authors(s) and do not necessarily reflect the views of the National Science Foundation.  The funders had no role in study design, data collection and analysis, decision to publish or preparation of the manuscript.
AMB was supported by the H.\ Mason Keeler Endowed Professorship in Sports Fisheries Management.
Special thanks to Colin Torney for providing the code to make the score field gradients, to Hongbo Fang for assisting with analyses and to Rob Goldstone for helpful feedback on the interpretation of our results.
 
\section*{Author Contributions}

\small

RH, PK, AP, NG, and JT formulated  the study. RH and PK designed the experiments, implemented the experiments, and analyzed the data. RH, PK, and AB wrote the manuscript. All authors gave final approval for publication and agree to be held accountable for the work performed therein.

\section*{Competing Interests}

\small

The authors declare no competing interests.

\renewcommand{\thesection}{S\arabic{section}}
\renewcommand{\thefigure}{S\arabic{figure}}

\setcounter{section}{0}
\setcounter{figure}{0}
\setcounter{table}{0}

\section*{Timeline}

\begin{itemize}
    \item Exp.~1 was run between Oct. 26 and Nov. 2, 2016.
    \item Exp.~2 was run between Nov. 20 and Nov. 28, 2017.
    \item Exp.~3 was run between Jan. 25 and Jan. 31, 2015.
\end{itemize}

\section*{Supplementary Videos}

Reconstructed videos of all human group sessions are available at \url{https://github.com/hawkrobe/emergent-sensing}, along with example videos of simulated groups under different models.

%\addcontentsline{toc}{chapter}{\listfigurename}
%\listoffigures

\begin{figure}
  \centering
  \includegraphics[width=0.45\textwidth]{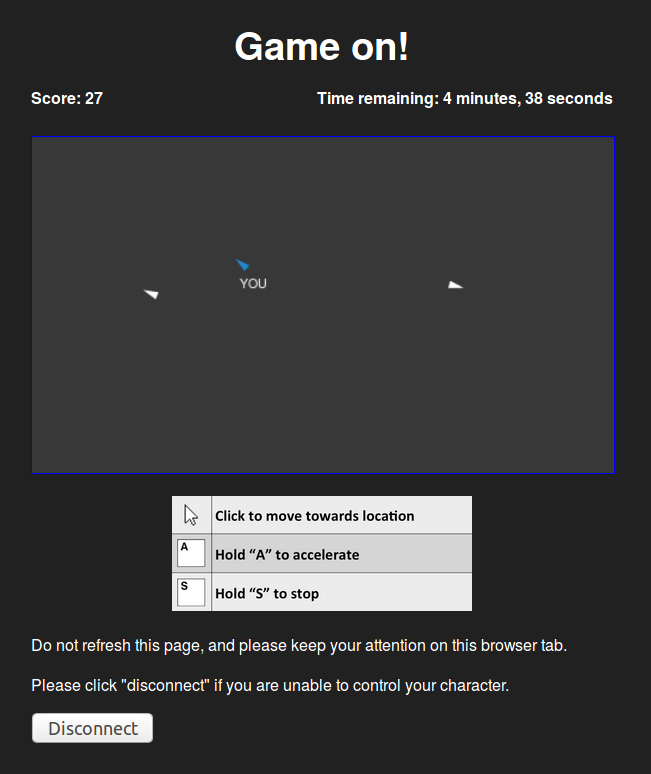}
\includegraphics[width=0.45\textwidth]{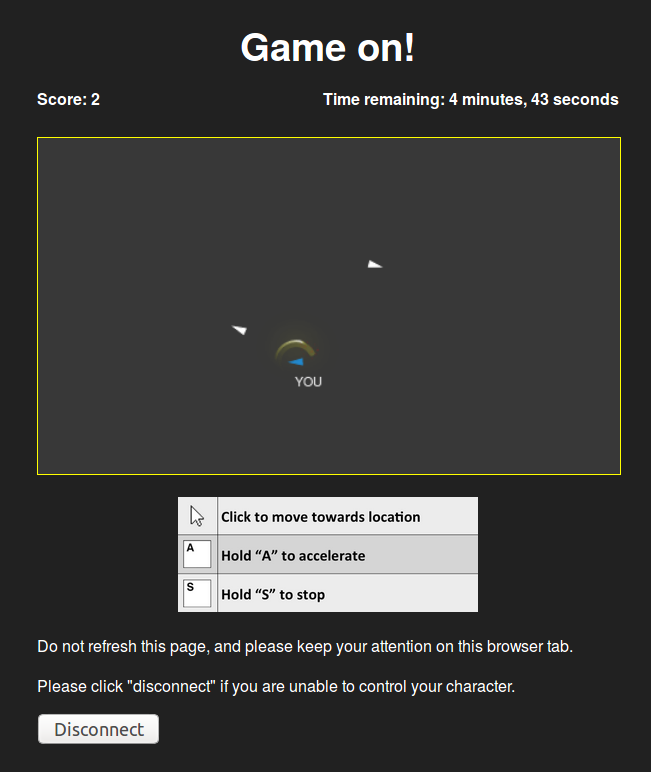}
  \includegraphics[width=0.5\textwidth]{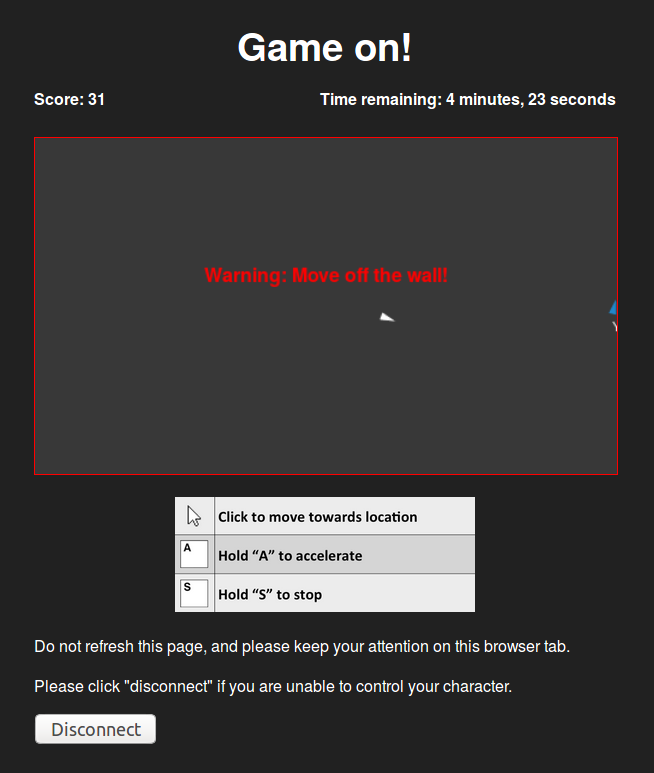}
  \hspace{0.1cm}
  \caption{Examples of Experiment 1 interface. The score shows the total accumulated rewards the participant has received up to the current time.}
  \label{fig:supplemental_interface}
\end{figure}

\begin{figure}
  \centering
  \includegraphics[width=0.99\textwidth]{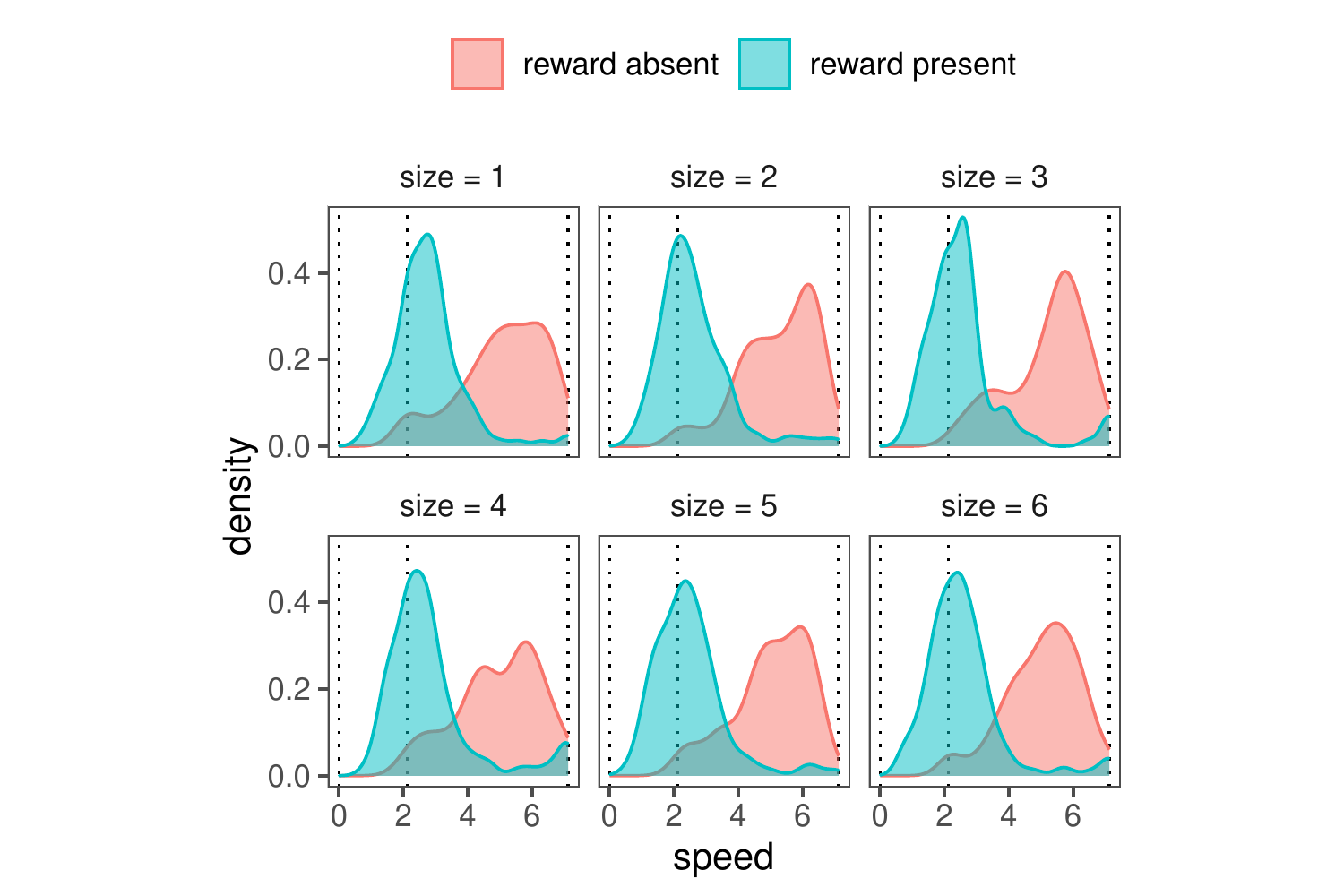}
  \hspace{0.1cm}
  \caption{Participants move faster on average when reward is absent (red) than when reward is present (blue) in Experiment 1. Speed distributions are similar across group sizes (panels). Vertical lines represent the three speeds available to participants in the user interface, with the center line being the default speed.}
  \label{fig:speeds}
\end{figure}

\begin{figure*}
\centering
  \includegraphics[width=0.45\textwidth,trim=2.5cm 3cm 2cm 3cm,clip]{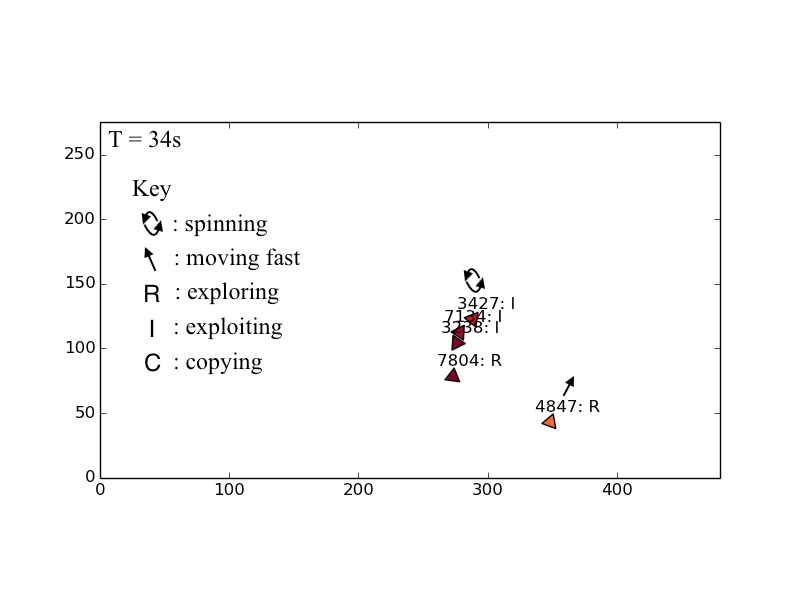}
  \includegraphics[width=0.45\textwidth,trim=2.5cm 3cm 2cm 3cm,clip]{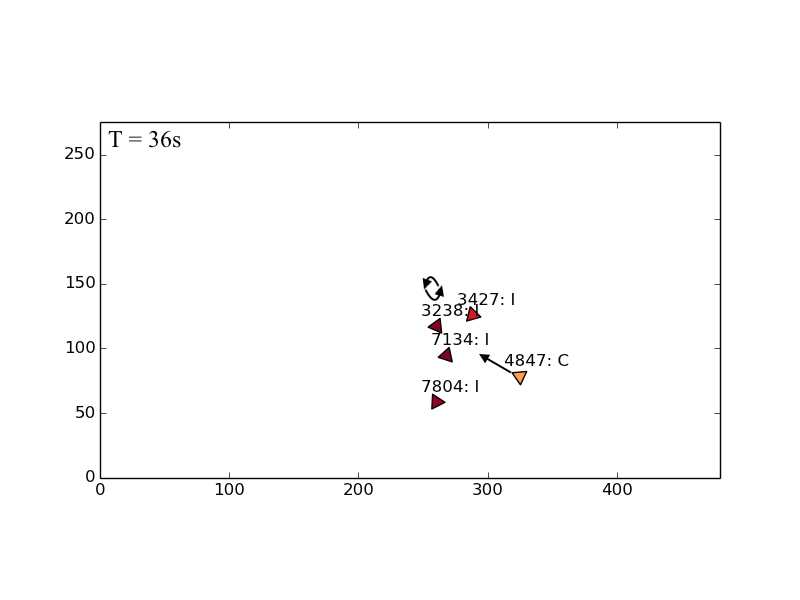}
  \includegraphics[width=0.45\textwidth,trim=2.5cm 3cm 2cm 3cm,clip]{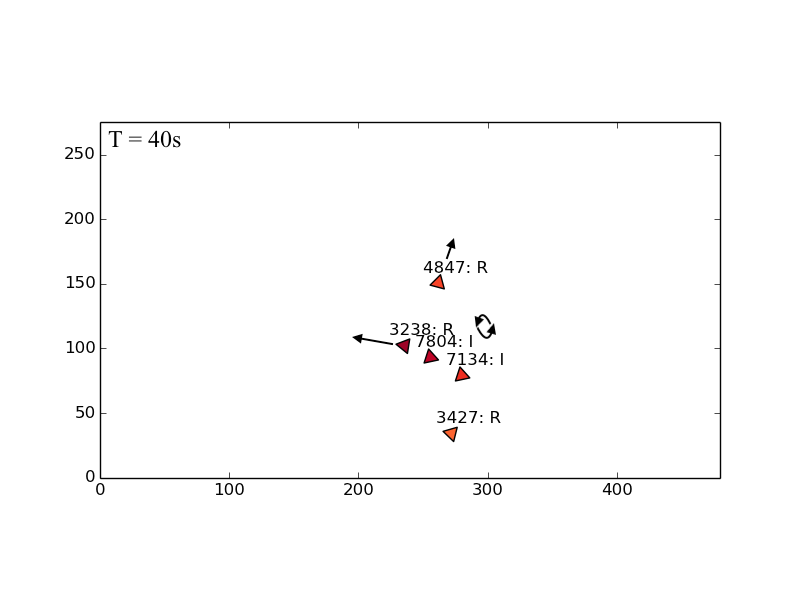}
  \includegraphics[width=0.45\textwidth,trim=2.5cm 3cm 2cm 3cm,clip]{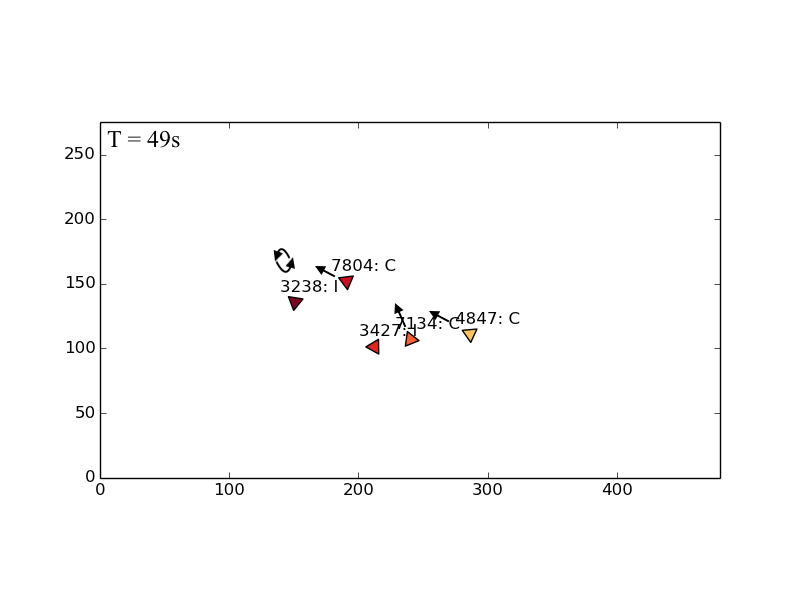} % 388
  \includegraphics[width=0.45\textwidth,trim=2.5cm 3cm 2cm 3cm,clip]{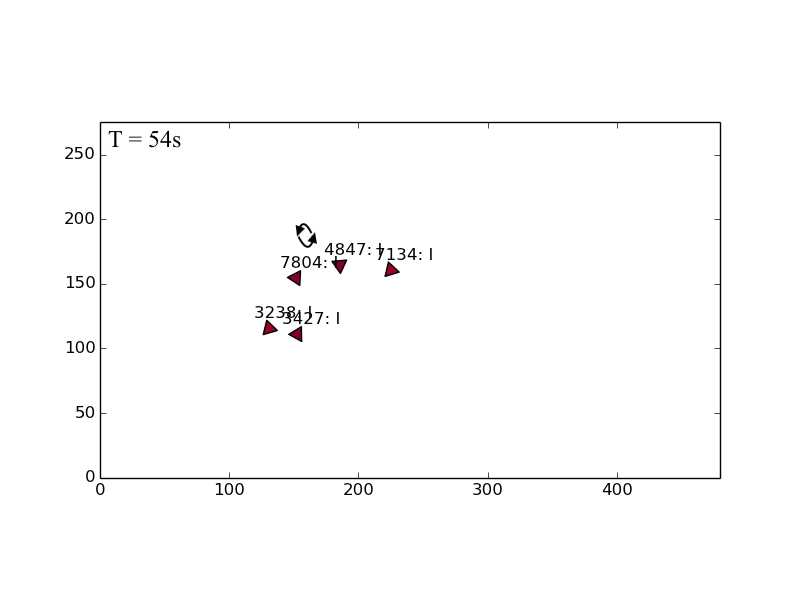}
  \includegraphics[width=0.45\textwidth,trim=2.5cm 3cm 2cm 3cm,clip]{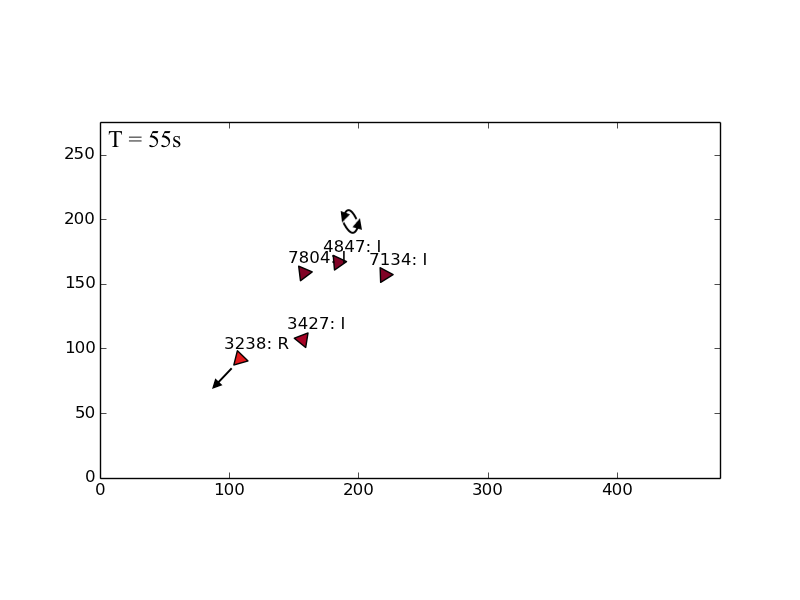}
  \caption{Reconstructions of actual gameplay in a five-person group in Experiment 3,  illustrating both failed exploration leading to selective copying and successful exploration leading to collective
    movement. Colors indicate the individuals' scores, with red being
    higher and orange/yellow being lower.  The participant labels indicate
    both participant IDs and also the participant states our feature extraction
    procedure inferred.  Other annotations are provided to give a
    sense for the game dynamics.  At $34$ seconds, in the first panel,
    most of the group has converged on exploiting a particular area
    while one individual is exploring independently.  To the right, at
    36 seconds, the exploring individual appears to have failed to
    find a good location and ceases exploring by copying the group.
    At 40 seconds, the final panel in the first row, the score field
    has shifted and some of the group begins exploring while others
    continue to exploit.  By 49 seconds, the first panel in the second
    row, one of the exploring individuals found a good location, and
    other participants have begun to move towards that individual.  At 54
    seconds, the entire group is exploiting the new area.  In the
    final panel, at 55 seconds, the background has shifted enough
    again that one of the individuals begins to explore.}
  \label{fig:example}
\end{figure*}

\begin{figure}[b!]
  \centering
  \includegraphics[width=0.4\textwidth]{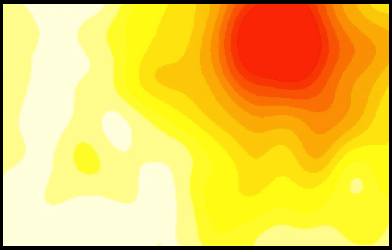}
  \hspace{0.1cm}
  \includegraphics[width=0.4\textwidth]{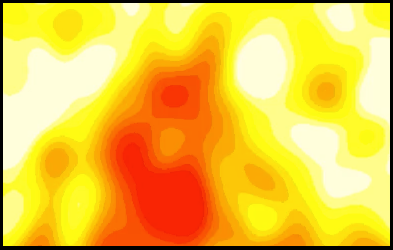}
  \caption{Example snapshots of score fields from the low noise (left) and high noise (right) conditions used in Experiment 3.  Red areas indicate higher scoring areas.}
  \label{fig:score_exp3}
\end{figure}

\begin{figure}
  \centering
  \includegraphics[width=0.9\textwidth]{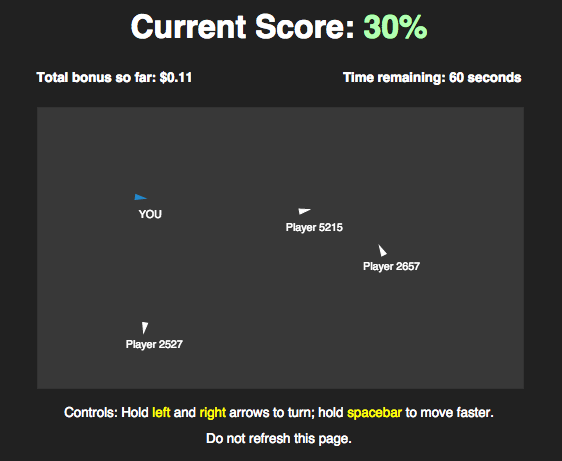}
  \caption{Screenshot of the Experiment 3 interface.  The
    score displayed corresponds to the value of the score field at the
    location that the participant's avatar is occupying.}
  \label{fig:exp3_interface}
\end{figure}

% \begin{figure}
%   \centering
%   \includegraphics[width=0.9\textwidth]{./figures/FigS6}
%   \caption{Auto-correlations in performance; after approximately 100 ticks of the physics engine (12.5 seconds), an individual's score is effectively uncorrelated, on average.}
%   \label{fig:exp3_interface}
% \end{figure}

\end{document}